# Localization of chiral edge states by the non-Hermitian skin effect

Gui-Geng Liu[1,#], Subhaskar Mandal[1,#], Peiheng Zhou[2,#], Xiang Xi[3,#], Rimi Banerjee[1], Yuan-Hang Hu[2], Minggui Wei[1], Maoren Wang[2], Qiang Wang[4], Zhen Gao[3], Hongsheng Chen[5], Yihao Yang[5,*], Yidong Chong[1,6,*], and Baile Zhang[1,6,*]

[1]Division of Physics and Applied Physics, School of Physical and Mathematical Sciences, Nanyang Technological University, 21 Nanyang Link, Singapore 637371, Singapore.

[2]National Engineering Research Center of Electromagnetic Radiation Control Materials, State Key Laboratory of Electronic Thin Film and Integrated Devices, University of Electronic Science and Technology of China, Chengdu 610054, China.

[3]Department of Electrical and Electronic Engineering, Southern University of Science and Technology, Shenzhen 518055, China.

[4]School of Physics, Collaborative Innovation Center of Advanced Microstructures, Nanjing University, Nanjing, Jiangsu 210093, China.

[5]Interdisciplinary Center for Quantum Information, State Key Laboratory of Modern Optical Instrumentation, ZJU-Hangzhou Global Science and Technology Innovation Center, College of Information Science and Electronic Engineering, Key Lab. of Advanced Micro/Nano Electronic Devices & Smart Systems of Zhejiang, ZJU-UIUC Institute, Zhejiang University, Hangzhou 310027, China.

[6]Centre for Disruptive Photonic Technologies, The Photonics Institute, Nanyang Technological University, 50 Nanyang Avenue, Singapore 639798, Singapore.

[#]These authors contributed equally to this work.

[*]E-mail: *yangyihao@zju.edu.cn (Y.Y.); yidong@ntu.edu.sg (Y.C.); blzhang@ntu.edu.sg (B.Z.)*

**Abstract**

Quantum Hall systems host chiral edge states extending along the one-dimensional boundary of any two-dimensional sample. In solid state materials, the edge states serve as perfectly robust transport channels that produce a quantised Hall conductance; due to their chirality, and the topological protection by the Chern number of the bulk bandstructure, they cannot be spatially localized by defects or disorder. Here, we show experimentally that the chiral edge states of a lossy quantum Hall system can be localized. In a gyromagnetic photonic crystal exhibiting the quantum Hall topological phase, an appropriately structured loss configuration imparts the edge states' complex energy spectrum with a feature known as point-gap winding. This intrinsically non-Hermitian topological invariant is distinct from the Chern number invariant of the bulk (which remains intact) and induces mode localization via the "non-Hermitian skin effect". The interplay of the two topological phenomena – the Chern number and point-gap winding – gives rise to a non-Hermitian generalisation of the paradigmatic Chern-type bulk-boundary correspondence principle. Compared to previous realisations of the non-Hermitian skin effect, the skin modes in this system have superior robustness against local defects and disorders.

**Introduction**

The significance of band topology was first discovered in the context of the quantum Hall effect in the 1980s[1]. In certain two-dimensional (2D) materials with broken time-reversal symmetry, the Hall conductance is exactly quantised, which is intimately tied to the fact that the bulk is insulating, and charge transport occurs exclusively via chiral edge states[2]. The existence of the chiral edge states is guaranteed by the bulk bands' nontrivial topology, as characterized by a topological invariant (the Chern number $C$[3-5]; see Fig. 1a). It is important for the quantum Hall effect that the chiral edge states are immune to backscattering (so long as the bulk is insulating), and hence do not undergo Anderson localization despite being one-dimensional (1D) transport channels. The robustness of chiral edge states against localization has even been directly observed in classical wave realisations of quantum Hall systems, based on photonic[6,7], acoustic[8,9], and mechanical lattices[10].

Theories of band topology have mostly been developed in the context of quantum Hall phases, and other topological phases of matter that are all Hermitian (energy-conserving). In recent years, however, there has been a great deal of progress in understanding the topology of non-Hermitian materials[11-16]. Some concepts from Hermitian band topology, including the Chern number[12], turn out to be generalizable to the non-Hermitian regime. More interestingly, non-Hermitian systems have been found to possess unique forms of band topology with no counterpart in the Hermitian regime. For instance, non-Hermitian band spectra may exhibit "point gaps," defined by points in the complex plane that are encircled by but do not overlap with any band frequency (energy)[13,15]. (This is distinct from a "line gap", an arbitrary line in the complex plane separating the bands, which is the more straightforward generalization of the Hermitian notion of a band gap.) A point-gapped non-Hermitian Hamiltonian cannot be continuously deformed to a Hermitian Hamiltonian without closing the point gap. Moreover, a point gap can be associated with an integer winding number, and nonzero windings are associated with the "non-Hermitian skin effect" (NHSE)[17-47], whereby an extensive number of bulk states become localized to a boundary. This phenomenon has been observed in recent experiments based on classical-wave systems[37-47]. To our knowledge, the implications of the NHSE for chiral edge states, which involve the interplay between point-gap topology and Chern number topology[34,35], have never been studied in any experiment.

In this work, we report on the observation of NHSE-induced localization of chiral edge states in a gyromagnetic photonic crystal (PhC). Such PhCs are commonly known as photonic topological insulators, and exhibit photonic bandgaps with nonzero Chern numbers[6]. By intentionally introducing losses to the system (thereby making it non-Hermitian) in a particular spatial pattern, we induce point gaps in the complex spectrum of the chiral edge states, with a pair of winding numbers ($v_x$, $v_y$) associated with the $x$ and $y$ directions respectively (see Fig. 1b). We show that the point-gap winding topology determines whether the chiral edge states are localized along a 1D sample edge (when $v_x \neq 0$ or $v_y \neq 0$), or at a zero-dimensional (0D) corner (when both $v_x \neq 0$ and $v_y \neq 0$). The chiral edge states are thus governed by a hybrid invariant ($C$; $v_x$, $v_y$) involving both point-gap and Chern number topology (see Extended Data Fig. 1 for an illustration of the hybrid bulk-boundary correspondence).

Our system also behaves differently from other NHSE realisations that are not based on chiral edge states[36-47]. Most of the theoretical models exhibiting the NHSE, such as the Hatano-Nelson model[13], are based on discrete (tight-binding) lattices, and the most common route to manifesting the NHSE is to introduce asymmetric couplings between discrete lattice sites, which have different magnitudes in the forward and backward directions (this determines the direction in which waves are "funneled" [37], and consequently where the skin modes are localized[16]). Our gyromagnetic PhC is continuous rather than discrete, and it does not require asymmetric couplings, which have been challenging to implement on many experimental platforms. Moreover, because the skin effect in this system is based on chiral edge states, forward and backward transport is spatially separated (on opposite edges of a strip), which provides the skin effect with superior robustness against local defects and disorder.

The PhC hosting localized chiral edge modes is depicted in Fig. 2a. It consists of a square lattice with three gyromagnetic rods per unit cell, with each rod surrounded by microwave-absorbing materials with a relative permittivity of either $\varepsilon_1 = \varepsilon_1' - i\varepsilon_1''$ or $\varepsilon_2 = \varepsilon_2' - i\varepsilon_2''$, where $\varepsilon_1' \approx \varepsilon_2'$ and $\varepsilon_1'' \ll \varepsilon_2''$ (see Methods). In the limit $\varepsilon_1'' = \varepsilon_2'' = 0$ and without losses of gyromagnetic rods, the PhC would be Hermitian and all eigenfrequencies would be real. The unit cell is mirror symmetric along the $x$ direction, but not along $y$. We apply a static magnetic field of 0.7 T along the $z$-axis, breaking time-reversal symmetry in the gyromagnetic rods (see Methods). The PhC is placed in a parallel-plate waveguide, and we focus on the transverse magnetic (TM) modes, which have electric fields polarized along the *z*-axis.

Due to the material losses, the PhC is non-Hermitian. Its bulk band spectrum, calculated from a single unit cell under periodic boundary conditions (PBCs) along both *x* and *y* ("double-PBCs"), forms a set of distinct complex bands each filling a bounded region of the complex frequency plane, as indicated by the blue areas in Fig. 2b. The bands are thus separated by line gaps. For the second and third gaps, depicted in this plot, the non-Hermitian generalisation of the Chern number is well-defined[12] and yields $C_2 = -1$ and $C_3 = 1$ (see Methods).

Next, we consider a finite-size sample of the PhC, bounded by copper claddings along both the *x* and *y* directions, as shown in Fig. 2c. Because the claddings act as perfect reflectors at microwave frequencies, this configuration is equivalent to open boundary conditions (OBCs) in a tight-binding model (i.e., open truncation of the lattice) along both *x* and *y* ("double-OBCs"). The calculated complex eigenfrequencies are plotted as diamond markers in Fig. 2b. We observe that a large set of eigenfrequencies, corresponding to bulk modes, occupy the same areas as the bulk band frequencies. However, there are also eigenfrequencies spanning the bulk line gaps; the corresponding eigenfunctions are strongly localized along the sample's right edge (for gap 2) or left edge (for gap 3), as shown by the two exemplary states plotted in Fig. 2d.

The existence of these edge states, which are localized to part of the sample boundary, arises jointly from point-gap winding (an intrinsically non-Hermitian topological invariant) and the Chern number (a topological band invariant generalised from the Hermitian regime). To demonstrate the former, we study the supercell depicted in Fig. 2e, which has PBCs along *x* and OBCs in the *y* direction ("*x*-PBC/*y*-OBC"). The calculated eigenfrequencies, plotted in Fig. 2f, form a spectral loop in each gap. As the wavenumber $k_x$ sweeps through the 1D Brillouin zone of the supercell, the eigenfrequencies in the second (third) gap advance anticlockwise (clockwise), producing a point-gap winding number of +1 (-1). This gives rise to the NHSE[25] and the localization behavior in the double-OBCs sample discussed in the previous paragraph.

Next, we plot intensity distributions for the supercell eigenstates in the bulk gap (Fig. 2g). The edge states are chiral: states of opposite $k_x$ lie on opposite (upper or lower) edges of the supercell, consistent with the bulk non-Hermitian Chern number of each gap. Moreover, the states on the upper (lower) arm of each spectral loop, whose eigenfrequencies have a larger (smaller) imaginary part, lie on the lower (upper) edge of the supercell. This can be understood intuitively from the supercell's configuration: the states on the lower edge (e.g., states 2 and 4 in Fig. 2g) travel through a lossier region than those on the upper edge (states 1 and 3 in Fig. 2g).

The point-gap winding of the PhC can be quantified using the winding numbers[25]

$$v_{\alpha,n} = \int_{BZ} \frac{dk_\alpha}{2\pi i} \frac{d}{dk_\alpha} \ln\left[f(k_\alpha) - f_0\right], \tag{1}$$

where $\alpha = x, y$ indicates the direction along which PBCs are imposed on a supercell, $n$ is the gap index, $f$ is the complex eigenfrequency, and $f_0$ is a reference frequency within the loop. As previously noted, $v_{x,2} = 1$ and $v_{x,3} = -1$; the sign difference between the two point-gap windings is consistent with the edge states being localized on opposite sides of the double-OBCs sample, as seen in Fig. 2d. Similarly, we can construct a supercell along $x$ ($y$-PBC/$x$-OBC), as depicted in Fig. 2h. In this case, we do not observe any spectral loops (Fig. 2i and j), so $v_{y,2} = 0$ and $v_{y,3} = 0$ and the NHSE does not occur. This is consistent with the lack of localization on the upper and lower edges in the double-OBCs system (Fig. 2d)[25].

To experimentally study the localization of the chiral edge states, we fabricate the sample depicted in Fig. 3a and b, and performe a pump-probe microwave measurement. We first place the source antenna at each of the four corners, in turn, and use a detection antenna to map out the excited electric fields (see Methods). The measured field distributions are plotted in Fig. 3c and d, for source frequencies of 15.3 GHz (in gap 2) and 17.2 GHz (in gap 3), respectively, with the source positions indicated by blue stars. As the PhC is lossy, the edge waves emitted by the source experience an overall exponential decay corresponding to mode profiles of $e^{-\beta_1 l}$, $e^{-\beta_2 l}$, $e^{-\beta_3 l}$, and $e^{-\beta_4 l}$ along the upper, right, lower, and left boundaries respectively, where $l$ is the path length along the boundary (measured clockwise and anticlockwise for gaps 2 and 3, respectively). To view the results more clearly, we counteract the overall decay by artificially amplifying the measured field distributions by an exponential gain factor $e^{\beta l}$, where $\beta = \frac{\beta_1 + \beta_2 + \beta_3 + \beta_4}{4}$ and $\beta_{1,2,3,4}$ are extracted by exponentially fitting the measured field distributions along each boundary. The precise distribution of this gain factor is specified in Extended Data Fig. 2. The compensated field distributions are plotted in Fig. 3e and f, revealing the accumulation of field intensity at the right (for gap 2) and left (for gap 3) sample edges, respectively. Notably, the localization of the compensated fields is independent of the source position. This measured behavior agrees with the eigenmode analysis (with complex eigenfrequencies) presented in Fig. 2d.

By adjusting the design of the PhC, we can obtain different point-gap winding numbers, so that the NHSE manifests differently. Figure 4a shows a PhC unit cell with broken mirror symmetry in both the *x* and *y* directions. The bulk spectrum (double-PBCs) is similar to the previous case, with clear line gaps. To determine the point-gap windings, we take supercells with *x*-PBC/*y*-OBC (Fig. 4c) and *y*-PBC/*x*-OBC (Fig. 4d); these two cases give identical spectra (Fig. 4b) containing loops in both gaps with $v_{x,2}=1$, $v_{y,2}=1$, $v_{x,3}=-1$, and $v_{y,3}=-1$. The NHSE is thus predicted to localize the chiral edge states towards both the +*x* and +*y* directions in gap 2, and towards the –*x* and –*y* directions in gap 3. We then construct a finite sample under double-OBCs (Fig. 4e) and plot its eigenfrequencies in Fig. 4f. In this system, gaps 2 and 3 are found to be spanned by edge states, whose eigenfunctions (Fig. 4g) are localized on the upper-right sample corner (for gap 2) and lower-left sample corner (for gap 3), respectively, consistent with predictions of the NHSE.

Next, we implement an experimental sample with the loss configuration described in the previous paragraph (see Fig. 4e). The electric field distributions are measured and analysed using the same method as before (see Extended Data Fig. 3 for the loss compensation factors). The results are plotted in Fig. 4h and i. The compensated electric field distributions are evidently concentrated at the sample's right-upper corner (Fig. 4h) and the left-lower corner (Fig. 4i) for gap 2 and gap 3, respectively. These results are consistent with the eigenmode profiles shown in Fig. 4g.

The above results can be understood in terms of a bulk-boundary correspondence governed by a hybrid topological invariant

$$v_n = (C_n; v_{x,n}, v_{y,n}) \qquad (2)$$

for each gap *n*. Only a combination of the Chern number and point-gap winding numbers can predict the existence and localization behavior of the edge states. The Chern number $C_n$ determines the number of chiral edge states in the gap, while the two point-gap winding numbers determine how they are localized in the *x* and *y* directions (see Extended Data Fig. 1 for details). In Extended Data Fig. 4, we study the boundary modes in a gap with a large Chern number $C_n = 2$ and find two branches of localized chiral edge states spanning in the gap.

Because the skin modes in this system are based on chiral edge states, they exhibit superior topological protection against local defects. Previous experimental demonstrations of the NHSE have relied on discrete lattices with asymmetric couplings (which can be thought of as funneling bulk states in the direction of the dominant coupling strength). In such designs, a sufficiently strong local defect can serve as a potential barrier that creates its own skin modes, preventing skin modes from appearing on the desired lattice boundary (see Extended Data Fig. 6g-h).

In the PhC, by contrast, the forward and backward transport of energy occurs via chiral edge states that are spatially separated at opposite edges of a strip sample. The chirality is protected by the nonzero Chern number, which cannot be overturned by local defects. Hence, a local defect along either edge does not localize skin modes, and all skin modes remain on the system boundary (see Extended Data Fig. 6a-f).

In conclusion, we have demonstrated the existence of edge and corner localized chiral edge states in a gyromagnetic PhC with NHSE. The Chern-type bulk-boundary correspondence has been generalized to the non-Hermitian regime, for which we have proposed a hybrid topological invariant that incorporates both the Chern number and two winding numbers to accurately predict the boundary states in both Hermitian and non-Hermitian cases. Furthermore, our work represents the first observation of the NHSE in a PhC, going beyond the tight binding approximation[32]. Due to the role of the chiral edge states, the NHSE in this system exhibits superior robustness against local defects, compared to previous NHSE realizations. Our findings may find applications in lasing[48] and robust light harvesting[37].

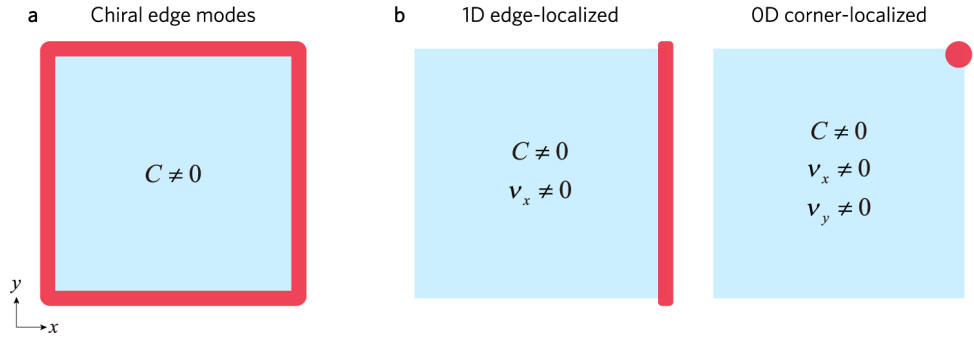

**Fig. 1 | Chern-type bulk-boundary correspondence and its non-Hermitian generalization. a,** Chiral edge states distribute over all edges in a two-dimensional system characterized by a Chern number $C \neq 0$. **b,** Non-Hermitian topology characterized by two winding numbers $\nu_x$ and $\nu_y$ can localize chiral edge states. Left: When either $\nu_x \neq 0$ or $\nu_y \neq 0$, chiral edge states are localized at an edge. Right: When both $\nu_x \neq 0$ and $\nu_y \neq 0$, chiral edge states are localized at a corner. The determination of boundary modes requires a hybrid topological invariant $(C; \nu_x, \nu_y)$.

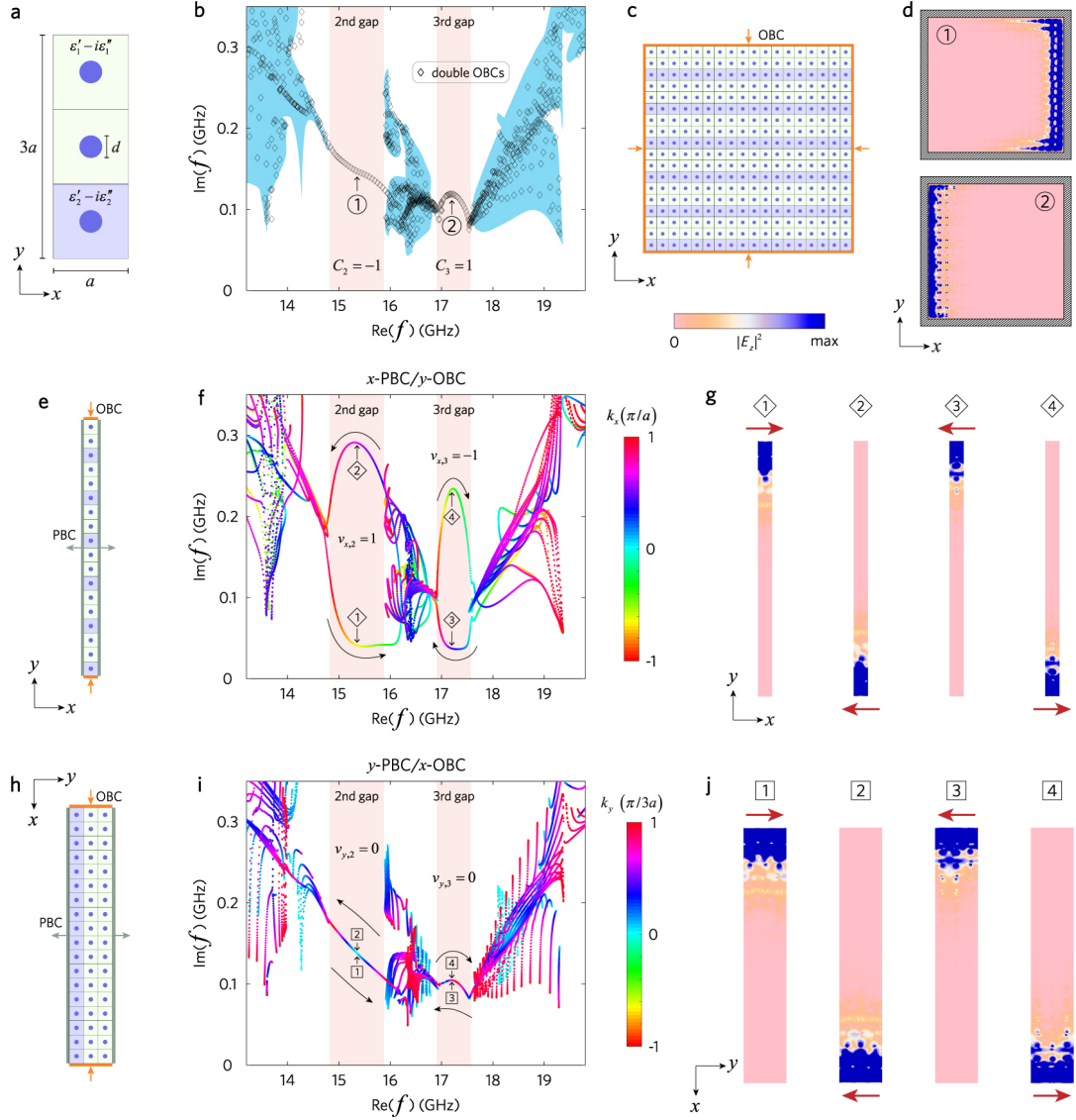

**Fig. 2 | First-principle study of edge-localized chiral edge states. a,** Unit cell of the PhC. The lattice constants along the $x$- and the $y$-direction are $a$ and $3a$, respectively, with $a = 10$ mm. The diameter of the gyromagnetic rods is $d = 3$ mm. **b,** Eigenfrequencies for double PBCs (blue region) and double OBCs (rhombic dots). **c,** Finite structure with double OBCs. **d,** Eigenfunctions for states labelled "1" and "2" in **b**. **e,** Supercell with $x$-PBC/$y$-OBC. **f,g,** Eigenfrequencies and eigenfunctions for the supercell in **e**. The black arrows in **f** donate the $k_x$ increasing directions. The red arrows in **g** indicate the group velocities of the edge states. **h-j,** Similar to **e-g** but for $y$-PBC/$x$-OBC.

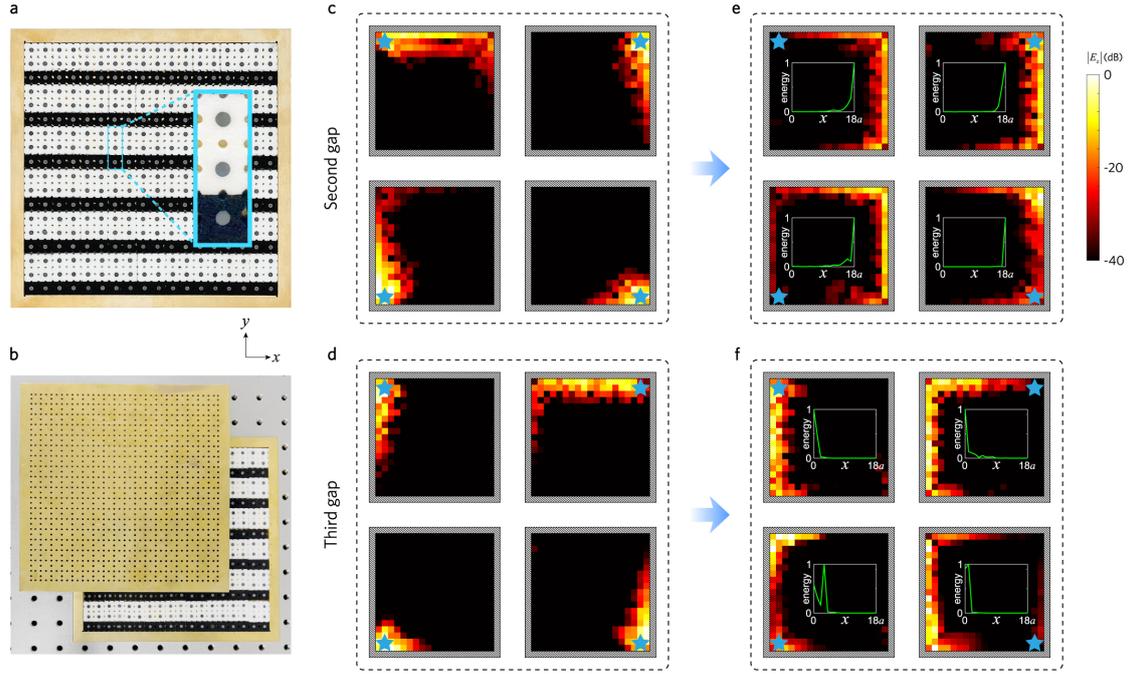

**Fig. 3 | Observation of edge-localized chiral edge states. a,b,** Photograph of the fabricated PhC. The top copper plate is removed (**a**) and shifted (**b**) for clear visualization. The inset in (**a**) shows an enlarged unit cell. The top copper and the background materials are drilled with small holes to facilitate experimental measurements. **c,d,** Original measured electric fields excited by a source antenna (denoted as blue stars) with frequencies at 15.3 GHz in the second gap (**c**) and at 17.2 GHz in the third gap (**d**), respectively. **e,f,** Similar to **c,d** but with artificial amplification. The electric fields are normalized within each panel itself. Insets in (**e**) and (**f**): average intensities over $y$-direction versus $x$-axis.

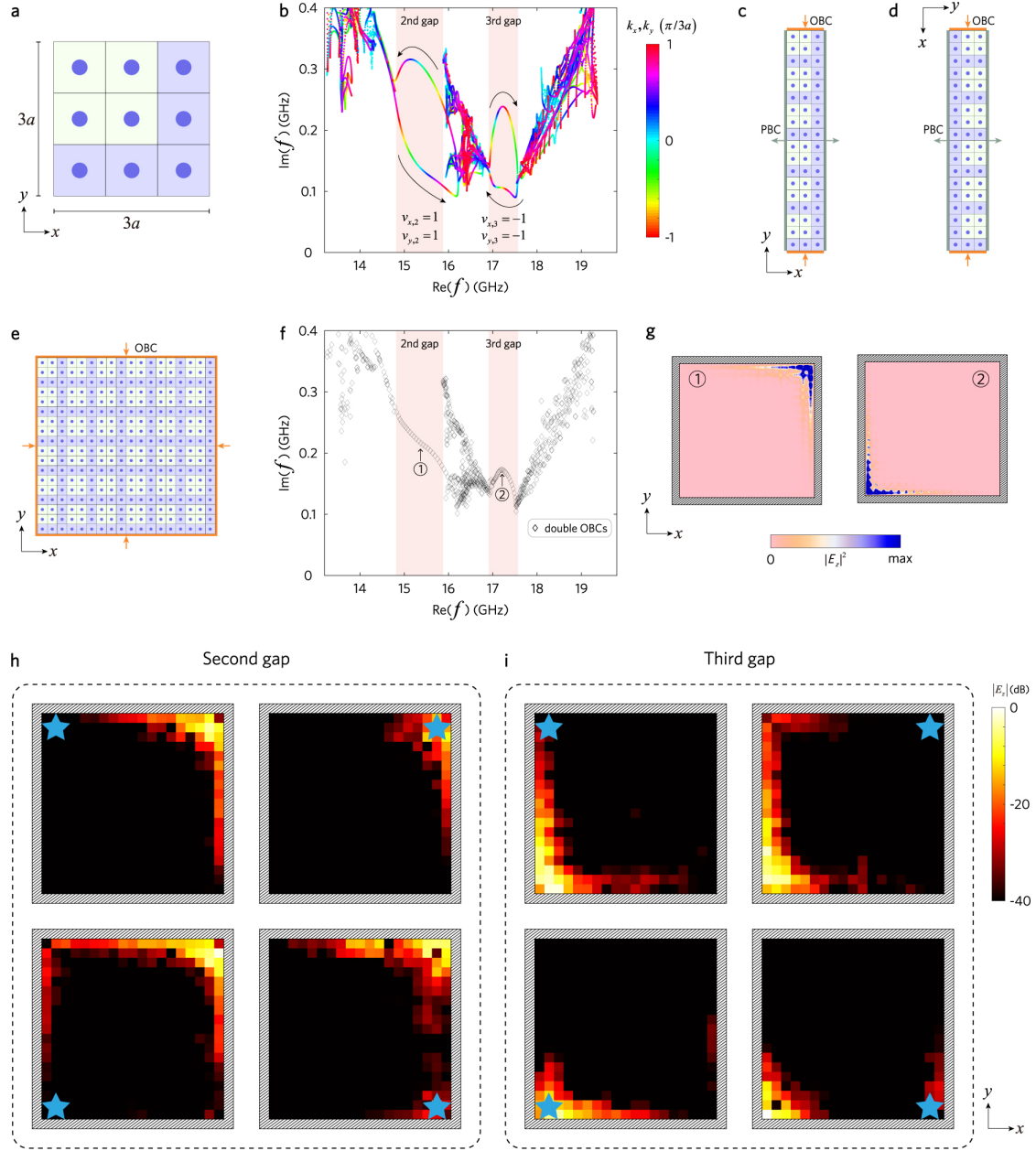

**Fig. 4 | Observation of corner-localized chiral edge states. a,** Unit cell of the PhC. **b,** Eigenfrequencies for the supercell in (**c**) under *x*-PBC/*y*-OBC and (**d**) under *y*-PBC/*x*-OBC. **e,** Finite structure with double OBCs. **f,g,** eigenfrequencies and eigenfunctions of the sample in (**e**). **h,i,** Measured electric fields with artificial amplification excited by a source antenna (denoted as blue stars) at 15.3 GHz in the second gap (**h**) and at 17.2 GHz in the third gap (**i**), respectively. The electric fields are normalized within each panel itself.